\documentclass[10pt,aps,twocolumn,prd,superscriptaddress,noshowpacs,nofootinbib,noshowkeys,floatfix]{revtex4}
\usepackage[dvips]{graphics,graphicx}
\usepackage[colorlinks=true,linktocpage=true,linkcolor=blue,citecolor=blue]{hyperref}
\usepackage[usenames,dvipsnames]{color}
\usepackage{amsmath, amssymb}
\usepackage{multirow}
\usepackage{longtable}
\usepackage{color}
\usepackage[normalem]{ulem}  

\renewcommand\sout{\bgroup \color{blue} \ULdepth=-.5ex \ULset}

\begin{document}

\preprint{}

\title{Weak and strong coupling limits of the Boltzmann equation in the relaxation-time approximation}

\author{Amaresh Jaiswal}
\affiliation{GSI, Helmholtzzentrum f\"ur Schwerionenforschung, Planckstrasse 1, D-64291 Darmstadt, Germany}
\author{Bengt Friman}
\affiliation{GSI, Helmholtzzentrum f\"ur Schwerionenforschung, Planckstrasse 1, D-64291 Darmstadt, Germany}
\author{Krzysztof Redlich}
\affiliation{Institute of Theoretical Physics, University of Wroclaw, PL-50204 Wroclaw, Poland}
\affiliation{Department of Physics, Duke University, Durham,  North Carolina 27708, USA}
\affiliation{Extreme Matter Institute EMMI, GSI, Planckstrasse 1, D-64291 Darmstadt, Germany}

\date{\today}

\begin{abstract}

We consider a momentum dependent relaxation time for the Boltzmann 
equation in the relaxation time approximation. We employ a power law 
parametrization for the momentum dependence of the relaxation time, 
and calculate the shear and bulk viscosity, as well as, the charge 
and heat conductivity. We show, that for the two popular 
parametrizations, referred to as  the linear and quadratic ansatz, 
one can obtain transport coefficients which corresponds to the weak 
and strong coupling regimes, respectively. We also show that, for a 
system of massless particles with vanishing chemical potential, the 
off-equilibrium corrections to the phase-space distribution function 
calculated with the quadratic ansatz are identical with those of the 
Grad's 14-moment method.

\end{abstract}



\maketitle


\section{Introduction}

In ultra-relativistic heavy-ion collisions, the energy density in 
the initial state can exceeds the critical value, predicted by 
Lattice QCD, for the existence of the hadronic matter \cite{domein}. 
At such conditions, quarks and gluons are deconfined and form a new 
state of matter called ``quark-gluon plasma" (QGP). It is now well 
established, that the QGP is indeed formed in nucleus-nucleus 
collisions, already at energies accessible at the BNL Relativistic 
Heavy Ion Collider (RHIC) \cite{Adams:2005dq, Adcox:2004mh} and the 
CERN Large Hadron Collider (LHC) \cite{ALICE:2011ab, ATLAS:2012at, 
Chatrchyan:2013kba}. It is also confirmed experimentally, that the 
QGP behaves as a nearly perfect fluid with a very small shear 
viscosity-to-entropy density ratio, $\eta/s$ \cite 
{Romatschke:2007mq, Song:2007ux, Luzum:2009sb, Song:2010mg, 
Schenke:2010rr, Bhalerao:2015iya, Jaiswal:2015saa}. Consequently, 
relativistic dissipative hydrodynamics has been quite successful in 
describing the space-time evolution of the QGP and its transport 
properties \cite{Heinz:2013th}.

The Boltzmann equation has been used to derive the dissipative 
hydrodynamic equations \cite {Grad, Muller:1967zza, Chapman, 
Anderson_Witting, Israel:1979wp, Prakash:1993bt, Dusling:2009df, 
Denicol:2010xn, Denicol:2012cn, Romatschke:2011qp, Jaiswal:2013fc, 
Jaiswal:2012qm, Jaiswal:2013npa, Jaiswal:2013vta, 
Chattopadhyay:2014lya, Jaiswal:2014isa, Florkowski:2015lra, 
Bhalerao:2013aha, Bhalerao:2013pza, Jaiswal:2015mxa, 
Romatschke:2003ms, Martinez:2012tu, Florkowski:2013lza, 
Bazow:2013ifa, Tsumura:2013uma, Tsumura:2015fxa, Kikuchi:2015swa}. 
It is a transport equation which governs the space-time evolution of 
the single particle phase-space distribution function, and is 
capable to accurately describe the microscopic dynamics of a system 
in the dilute limit. Moreover, in the limit of small mean free path, 
the Boltzmann equation starts to describe hydrodynamics. Therefore, 
derivation of the equations of dissipative hydrodynamics and its 
associated transport coefficients from the Boltzmann equation, is of 
importance, to characterize the non-equilibrium dynamics of a system.

Despite its advantages, the Boltzmann equation is difficult to solve 
directly because its collision integral depends on the product of 
the distribution functions. Simpler approximations for the collision 
term have been proposed, of which the relaxation-time approximation 
by Anderson and Witting, is the most commonly used model \cite
{Anderson_Witting}. The relaxation-time approximation for the 
collision term assumes, that the collisions between particles tend 
to restore the distribution function to its local equilibrium value, 
exponentially. This is an excellent approximation when the system is 
close to local thermodynamic equilibrium.

In the Anderson-Witting model, the Boltzmann relaxation time is 
assumed to be independent of the particle momenta. However, in 
general, the relaxation time can be momentum dependent and might 
show different functional dependence for different theories \cite 
{Dusling:2009df}. In this paper, we consider a power law 
parametrization for the momentum dependence of the relaxation time 
of the Boltzmann equation. We derive expressions for transport 
coefficients, such as shear and bulk viscosity, as well as, charge 
and heat conductivity. We show, that for two popular 
parametrizations, referred to as the linear and quadratic ansatz, 
the first viscous correction to the distribution function leads to 
identical expressions as that obtained using the Chapman-Enskog, and 
Grad's 14-moment method, respectively. We also demonstrate that the 
ratios of transport coefficients in these two cases corresponds to 
the weak and strong coupling regimes.


\section{Relativistic hydrodynamics}

The conserved energy-momentum tensor and particle four-current can
be expressed in terms of the single particle phase-space
distribution function $f(x,p)$ \cite{deGroot},  as
\begin{align}
T^{\mu\nu} &= \int dp \ p^\mu p^\nu (f+\bar f) = \epsilon u^\mu u^\nu-(P+\Pi)\Delta ^{\mu \nu}
+ \pi^{\mu\nu} , \label{en_mom_ten}\\
N^\mu &= \int dp \ p^\mu (f-\bar f) = nu^\mu + n^\mu . \label{cons_curr}
\end{align}
Here $dp=gd{\bf p}/[(2 \pi)^3\sqrt{{\bf p}^2+m^2}]$, $g$ and $m$ are 
the degeneracy factor and particle rest mass, $p^{\mu}$ is the 
particle four-momentum, and $f$ and $\bar f$ are the phase-space 
distribution functions for particles and anti-particles, 
respectively. Here we consider a system consisting of a single 
species of particles. In the tensor decompositions, $\epsilon$, $P$ 
and $n$ are the energy density, pressure and net number density, 
respectively, and $\Delta^{\mu\nu}=g^{\mu\nu}-u^\mu u^\nu$ is the 
projection operator orthogonal to the hydrodynamic four-velocity 
$u^\mu$ defined in the Landau frame: $T^{\mu\nu} u_\nu=\epsilon 
u^\mu$. The bulk viscous pressure $\Pi$, the shear stress tensor 
$\pi^{\mu\nu}$, and the charge diffusion current $n^\mu$, are 
dissipative quantities. We work with the Minkowskian metric tensor 
$g^{\mu\nu}\equiv\mathrm{diag}(+,-,-,-)$.

The fundamental conservation equations of energy-momentum
$\partial_\mu T^{\mu\nu} =0$, and particle current  $\partial_\mu
N^{\mu}=0$, yields the evolution equations for $\epsilon$, $u^\mu$
and $n$, as
\begin{align}
\dot\epsilon + (\epsilon+P+\Pi)\theta - \pi^{\mu\nu}\sigma_{\mu\nu} &= 0,  \label{evol_e}\\
(\epsilon+P+\Pi)\dot u^\alpha - \nabla^\alpha (P+\Pi) + \Delta^\alpha_\nu \partial_\mu \pi^{\mu\nu}  &= 0, \label{evol_u}\\
\dot n + n\theta + \partial_\mu n^{\mu} &=0. \label{evol_n}
\end{align}
Here we use the standard notation:  $\dot A\equiv u^\mu\partial_\mu A$
for the co-moving derivative,  $\nabla^\mu\equiv\Delta^{\mu\nu}
\partial_\nu$ for the space-like derivative, $\theta\equiv
\partial_\mu u^\mu$ for the expansion scalar,  and $\sigma^{\mu\nu}
\equiv\frac{1}{2}(\nabla^\mu u^\nu+\nabla^\nu u^\mu)-\frac{1}{3}
\theta\Delta^{\mu\nu}$ for the velocity stress tensor.

The equilibrium quantities such as the energy density, the
thermodynamic pressure and the net number density,  can be defined in
terms of the equilibrium distribution function,  as
\begin{align}
\epsilon_0 &\equiv u_\mu u_\nu T^{\mu\nu}_0 = u_\mu u_\nu
\int dp \, p^\mu p^\nu (f_0+\bar f_0), \label{def_en_den}\\
P_0 &\equiv -\frac{1}{3}\Delta_{\mu\nu}T^{\mu\nu}_0 = -\frac{1}{3}\Delta_{\mu\nu}
\int dp \, p^\mu p^\nu (f_0+\bar f_0), \label{def_press}\\
n_0 &\equiv u_\mu N^\mu_0 = u_\mu \int dp \, p^\mu (f_0-\bar f_0), \label{def_num_den}
\end{align}
where the suffix $"0"$ denotes the corresponding values in
equilibrium.

 In this work we consider a system of Boltzmann gas for
which the equilibrium distribution function is given by
\begin{align}
f_0 &= \exp(-\beta u\cdot p + \alpha), \label{eq_dist_fn_p}\\
\bar f_0 &= \exp(-\beta u\cdot p - \alpha). \label{eq_dist_fn_a}
\end{align}
Here $\beta\equiv1/T$ is the inverse temperature, $\alpha\equiv\mu/T$
is the ratio of chemical potential to temperature,  and $u \cdot
p\equiv u_\mu p^\mu$. For such a system, the integrals in Eqs.~(\ref
{def_en_den})-(\ref{def_num_den}) can be solved analytically,  to obtain
\begin{align}
\epsilon_0 &= g\frac{T^4z^2}{\pi^2}\left[3K_2(z)+zK_1(z)\right]\cosh(\alpha), \label{en_den}\\
P_0 &= g\frac{T^4z^2}{\pi^2}K_2(z)\cosh(\alpha), \label{press}\\
n_0 &= g\frac{T^3z^2}{\pi^2}K_2(z)\sinh(\alpha), \label{num_den}
\end{align}
where $z\equiv m/T$ is the ratio of the particle mass to temperature
and $K_n$ are the modified Bessel functions of the second kind.

For a dissipative system, the thermodynamic temperature and the
chemical potential is defined by the matching condition
$\epsilon=\epsilon_0$ and $n=n_0$. The Navier-Stokes expressions for
the dissipative quantities can be written in terms of the
first-order gradients,  as
\begin{align}
\pi^{\mu\nu} &= 2\eta\sigma^{\mu\nu}, \label{shear_NS}\\
\Pi &= -\zeta\theta, \label{bulk_NS}\\
n^\mu &= \kappa_n\nabla^\mu \alpha. \label{charge_NS}
\end{align}
Here the transport coefficients $\eta$, $\zeta$ and $\kappa_n$, 
denote the shear and bulk viscosity, and the charge conductivity, 
respectively.

It is well known, that the first-order relativistic Navier-Stokes 
theory suffers from acusality and instabilities. These issues are 
solved by considering second-order corrections to the dissipative 
equations. On the other hand, the form of the first-order transport 
coefficients are sensitive to the nature of the microscopic 
interactions, and can be used to distinguish between a weakly and 
strongly coupled field theory.

For a system close to local thermodynamic equilibrium, the 
phase-space distribution function can be decomposed into equilibrium 
and non-equilibrium parts, $f=f_0+\delta f$, where $|\delta 
f|/f_0\ll1$. Therefore, from Eqs.~(\ref{en_mom_ten}) and (\ref 
{cons_curr}), the shear stress tensor $\pi^{\mu\nu}$, the bulk 
viscous pressure $\Pi$, and the particle diffusion current $n^\mu$, 
can be expressed in terms of $\delta f$, as
\begin{align}
\pi^{\mu\nu} &= \Delta^{\mu\nu}_{\alpha\beta} \int dp \, p^\alpha p^\beta
\left(\delta f + \delta \bar f\right) ,\label{shear}\\
\Pi &= -\frac{\Delta_{\alpha\beta}}{3} \int dp \, p^\alpha p^\beta
\left(\delta f+\delta\bar f\right) , \label{bulk}\\
n^\mu &= \Delta^\mu_\alpha \int dp \, p^\alpha
\left(\delta f - \delta \bar f\right) , \label{ch_curr}
\end{align}
where $\Delta^{\mu\nu}_{\alpha\beta}\equiv \frac{1}{2} 
(\Delta^{\mu}_{\alpha}\Delta^{\nu}_{\beta} + 
\Delta^{\mu}_{\beta}\Delta^{\nu}_{\alpha}) - \frac{1}{3} 
\Delta^{\mu\nu}\Delta_{\alpha\beta}$ is a traceless symmetric 
projection operator which is orthogonal to $u_\mu$ and 
$\Delta_{\mu\nu}$. In the following, we derive the Navier-Stokes 
expressions for the dissipative quantities, by iteratively solving 
the Boltzmann equation in the relaxation-time approximation to 
obtain $\delta f$, up to first order in gradients.


\section{Relaxation-time approximation}

Within kinetic theory, the evolution of the phase-space distribution 
function is governed by the Boltzmann equation. In the dilute limit, 
the Boltzmann equation provides a complete description of the 
microscopic dynamics of a system. In the present work, we consider a 
simplified version of the Boltzmann equation, where the collision 
term is written in the relaxation-time approximation \cite 
{Anderson_Witting},
\begin{equation}\label{boltz_eq}
p^\mu\partial_\mu f =  -\frac{u\cdot p}{\tau_R}\left( f-f_0 \right).
\end{equation}
Here, $\tau_R$ is the relaxation time for the Boltzmann equation
which, in general, can be a function of space-time, as well as,
the particle momenta.

For different microscopic theories, $\tau_R$, can exhibit a distinct 
functional dependence on particle momenta \cite{Dusling:2009df}. 
Therefore, to obtain the correct functional dependence, one should 
in general, consider the details of the microscopic dynamics. In the 
present work, however, we parametrize the momentum dependence of the 
relaxation time, with the following power law,
\begin{equation}\label{rel_time_para}
\tau_R(x,p) = \tau_0(x)\left(\frac{u\cdot p}{T}\right)^a,
\end{equation}
and consider two limiting cases:
\begin{enumerate}
\item $a=0$ (linear ansatz).
\item $a=1$ (quadratic ansatz).
\end{enumerate}
Most microscopic theories lie between these two extreme limits \cite 
{Dusling:2009df}. 

In the following, we demonstrate, that the transport coefficients 
obtained by using the linear ansatz, corresponds to weakly coupled 
microscopic theories, whereas those obtained with the quadratic 
ansatz, corresponds to strongly coupled theories.

In order to obtain the transport coefficients from Eqs.~(\ref 
{shear})-(\ref{ch_curr}), one needs to calculate $\delta f$. To that 
end, we solve Eq.~(\ref{boltz_eq}) iteratively, by employing a 
Chapman-Enskog like expansion \cite{Romatschke:2011qp, 
Jaiswal:2013npa}. The first-order solution is obtained as
\begin{equation} \label{first_ord_corr}
\delta f_1 = -\frac{\tau_R}{u\cdot p} \, p^\mu \partial_\mu f_0,
\end{equation}
which translates to
\begin{align}
\delta f_L &= -\frac{\tau_0}{u\cdot p} \, p^\mu \partial_\mu f_0, \label{first_ord_corr_L}\\
\delta f_Q &= -\frac{\tau_0}{T} \, p^\mu \partial_\mu f_0, \label{first_ord_corr_Q}
\end{align}
for the linear and quadratic ansatz, respectively. In the next 
section, we employ the above results for the linear and quadratic 
ansatzes for $\delta f$, to obtain expressions for the relativistic 
Navier-Stokes equations by evaluating the integrals in Eqs.~(\ref 
{shear})-(\ref{ch_curr}).


\section{Dissipative equations}

The first theoretical formulations of relativistic dissipative
hydrodynamics were proposed by Eckart \cite{Eckart:1940zz} and
Landau-Lifshitz \cite{Landau}. These formulations were the
relativistic analogues of the Navier-Stokes theory and involved
first-order gradients. In the following,  we derive relativistic
Navier-Stokes equations for the dissipative quantities. We consider
three different scenarios:
\begin{enumerate}
\item Both bulk viscous pressure and dissipative charge current
vanishes but shear stress tensor remains non-zero. Mathematically
this amounts to setting $m=0$ and $\mu=0$.
\item The dissipative charge current vanishes but shear stress tensor
and bulk viscous pressure remains non-zero. This is equivalent to
$m\ne0$ and $\mu=0$.
\item The bulk viscous pressure vanishes but shear stress tensor and
dissipative charge current are non-vanishing. This translates to $m=0$
and $\mu\ne0$.
\end{enumerate}
In each of the above three cases, we obtain the relativistic 
Navier-Stokes equations by using both linear and quadratic ansatz 
for $\delta f$ and compare the results.


\subsection{Case 1: $m=0$, $\mu=0$}

Since the bulk viscous pressure is proportional to $m^2$,  and 
current conservation equation vanishes for $\mu=0$, thus the only 
non-vanishing dissipative quantity in this case is the shear stress 
tensor. In order to derive first-order expression for $\pi^{\mu\nu}$, 
we need to obtain the derivatives of $\alpha$ and $\beta$. 
Considering the $z\to 0$ and $\alpha\to 0$ limits of Eqs.~(\ref 
{en_den})-(\ref{num_den}) and substituting in 
Eqs.~(\ref{evol_e})-(\ref{evol_n}), one gets
\begin{align}
\dot\beta &= \frac{\beta}{3}\theta - \frac{\beta}{3(\epsilon+P)}
\pi^{\mu\nu}\sigma_{\mu\nu}, \label{beta_dot_sh}\\
\nabla^\alpha\beta &= -\beta\dot u^\alpha - \frac{\beta}{\epsilon+P}
\Delta^\alpha_\mu \partial_\nu \pi^{\mu\nu}. \label{beta_nab_sh}
\end{align}
Using the above relations, Eqs.~(\ref{first_ord_corr_L}) and (\ref
{first_ord_corr_Q}) can be written as
\begin{align}
\delta f_L &= \frac{\tau_0\beta}{u\cdot p}\, f_0 \; p^\mu p^\nu \sigma_{\mu\nu}, \label{corr_L_sh}\\
\delta f_Q &= \frac{\tau_0\beta}{T}\, f_0 \; p^\mu p^\nu \sigma_{\mu\nu}. \label{corr_Q_sh}
\end{align}
It is now apparent, that while the coefficient of $f_0$ in Eq.~(\ref
{corr_L_sh}) is linear in momenta, in Eq.~(\ref {corr_Q_sh}) it is
quadratic; hence the nomenclature.

The first-order expression for $\pi^{\mu\nu}$, in the case of linear 
and quadratic ansatzes,  can be obtained by substituting Eqs.~(\ref 
{corr_L_sh}) and (\ref{corr_Q_sh}) into Eq.~(\ref{shear}). We get 
the relativistic Navier-Stokes equation,
\begin{equation} \label{NS_sh}
\pi^{\mu\nu} = 2 \, \tau_0\beta_\pi \, \sigma^{\mu\nu},
\end{equation}
where
\begin{equation}\label{betapi_sh}
\beta_\pi = \left\{ \begin{array}{rl}
\displaystyle\frac{1}{5}\left(\epsilon+P\right) &  \qquad \mbox{ (linear ansatz), } \\
\displaystyle\epsilon+P & \qquad \mbox{ (quadratic ansatz).} \end{array}\right.
\end{equation}
Comparing Eqs.~(\ref{NS_sh}) and Eq.~(\ref{shear_NS}), one gets, 
$\eta=\tau_0\beta_\pi$.

The difference in the expressions of $\beta_\pi$ in Eq.~(\ref
{betapi_sh}), obtained for the two ansatzes on the momentum 
dependent relaxation time, has some interesting consequences. 
Indeed, using Eqs.~(\ref {NS_sh}) and (\ref{betapi_sh}), one can 
rewrite Eqs.~(\ref{corr_L_sh}) and (\ref{corr_Q_sh}), as
\begin{align}
\delta f_L &= \frac{5f_0}{2(\epsilon+P)(u\cdot p)T} \, p^\mu p^\nu \pi_{\mu\nu}, \label{L_sh}\\
\delta f_Q &= \frac{f_0}{2(\epsilon+P)T^2} \, p^\mu p^\nu \pi_{\mu\nu}. \label{Q_sh}
\end{align}
Consequently, as could be expected, the $\delta f$ in Eq.~(\ref 
{L_sh}) is the same as that obtained using the iterative 
Chapman-Enskog method \cite{Bhalerao:2013pza, 
Chattopadhyay:2014lya}. On the other hand, the $\delta f$ in 
Eq.~(\ref{Q_sh}) is identical to that of the Grad's 14-moment 
method \footnote{For a detailed comparison see Ref.~\cite 
{Bhalerao:2013pza}}. This is indeed a very interesting and rather 
unexpected result, indicating that with a suitable choice of the 
momentum dependence of the relaxation time, the iterative 
Chapman-Enskog method can reproduce the $\delta f$ obtained using 
the moment method. A detailed analysis of this finding is left for a 
future work.

From the  phenomenological perspective, it is interesting to note, 
that the experimental results for the Hanburry-Brown-Twiss (HBT) 
radii favor the linear, rather than quadratic, momentum dependence 
of the viscous correction to the distribution function \cite 
{Bhalerao:2013pza}. Moreover, the transport results for the 
anisotropic flow and transverse momentum spectra also show agreement 
with the linear ansatz \cite{Plumari:2015sia}.


\subsection{Case 2: $m\ne0$, $\mu=0$}

In this case, the non-vanishing dissipative quantities are the shear 
stress tensor and the bulk viscous pressure. Considering the 
$\alpha\to 0$ limits of Eqs.~(\ref {en_den})-(\ref{num_den}),  and 
substituting them into Eqs.~(\ref{evol_e})-(\ref{evol_n}), we get
\begin{align}
\dot\beta &= \frac{\beta(\epsilon+P)}{3\epsilon+(3+z^2)P}\theta
+ \frac{\beta(\Pi\theta - \pi^{\mu\nu}\sigma_{\mu\nu})}{3\epsilon+(3+z^2)P}, \label{beta_dot_blk}\\
\nabla^\alpha\beta &= -\beta\dot u^\alpha - \frac{\beta}{\epsilon+P}
\left(\Pi\dot u^\alpha - \nabla^\alpha\Pi + \Delta^\alpha_\nu\partial_\mu\pi^{\mu\nu}\right). \label{beta_nab_blk}
\end{align}
Using the above relations, Eqs.~(\ref{first_ord_corr_L}) and (\ref
{first_ord_corr_Q}) become
\begin{align}
\delta f_L &= \frac{\beta\tau_0}{u\cdot p}\, f_0\!\left[ \frac{1}{3}\left\{p^2-(1-3c_s^2)
(u\cdot p)^2\right\}\theta + p^\mu p^\nu\sigma_{\mu\nu} \right], \label{FOC_sh_blk_L}\\
\delta f_Q &= \frac{\beta\tau_0}{T}\, f_0\!\left[ \frac{1}{3}\left\{p^2-(1-3c_s^2)
(u\cdot p)^2\right\}\theta + p^\mu p^\nu\sigma_{\mu\nu} \right], \label{FOC_sh_blk_Q}
\end{align}
where $c_s^2\equiv dP/d\epsilon$ is the velocity of sound squared.

Substituting Eqs.~(\ref{FOC_sh_blk_L}) and (\ref{FOC_sh_blk_Q}) in 
Eqs.~(\ref{shear}) and (\ref{bulk}), one gets the following 
Navier-Stokes equations for the shear stress tensor and the bulk 
viscous pressure, 
\begin{align}
\pi^{\mu\nu} &= 2\, \tau_0\beta_\pi\, \sigma^{\mu\nu} , \label{FOE_sh}\\
\Pi &= -\tau_0\beta_\Pi\, \theta . \label{FOE_blk}
\end{align}
The first-order transport coefficient of the shear stress tensor 
for these two ansatzes, is then obtained in the following form
\begin{equation}\label{betapi_sh_blk}
\beta_\pi = \left\{ \begin{array}{rl}
\displaystyle\beta\, I_{42}^{(1)} &  \qquad \mbox{ (linear ansatz), } \\
\displaystyle\epsilon+P & \qquad \mbox{ (quadratic ansatz),} \end{array}\right.
\end{equation}
where
\begin{equation}\label{I421}
I_{42}^{(1)} = \frac{gT^5z^5}{30\pi^2}\left[\frac{1}{16}(K_5-7K_3+22K_1)-K_{i,1}\right].
\end{equation}
Here the $z$-dependence of $K_n$ is implicitly understood,  and the
function $K_{i,1}$ is defined by the integral
\begin{equation}\label{kin}
K_{i,1}(z) = \int_0^\infty \frac{d\theta}{\cosh\theta}\,\exp(-z\cosh\theta),
\end{equation}
which can be evaluated as a Taylor series expansion up to any given 
order in $z$.

The first-order transport coefficient of the bulk viscous pressure, 
for the two ansatzes, is obtained as
\begin{equation}\label{betaPi_sh_blk}
\beta_\Pi = \left\{ \begin{array}{rl}
\displaystyle\frac{5}{3}\beta_\pi - (\epsilon+P)c_s^2 &  \qquad \mbox{ (linear ansatz), } \\
\displaystyle 2(\epsilon+P)\left(\frac{1}{3}-c_s^2\right) & \qquad \mbox{ (quadratic ansatz).} \end{array}\right.
\end{equation}
Note, that a comparison of Eqs.~(\ref{FOE_sh}) and (\ref{FOE_blk}) 
with Eqs.~(\ref{shear_NS}) and (\ref{bulk_NS}) gives $\eta= 
\tau_0\beta_\pi$ and $\zeta=\tau_0\beta_\Pi$. Therefore, the ratio 
of the coefficient of the bulk viscosity to that of shear viscosity, 
$\zeta/\eta=\beta_\Pi/\beta_\pi$, is independent of $\tau_0$, and 
can be written, as
\begin{equation}\label{blk_sh_ratio}
\frac{\zeta}{\eta} = \left\{ \begin{array}{rl}
\displaystyle 75\left(\frac{1}{3}-c_s^2\right)^2 &  \qquad \mbox{ (linear ansatz), } \\
\displaystyle 2\left(\frac{1}{3}-c_s^2\right) & \qquad \mbox{ (quadratic ansatz).} \end{array}\right.
\end{equation}
The expression for the linear ansatz is obtained by considering a 
small-$z$ expansion up to ${\cal O}(z^4)$. On the other hand, the 
result for the quadratic ansatz is exact.

The difference in the functional dependence of $\zeta/\eta$ on the 
sound velocity, obtained for the linear and quadratic ansatzes in 
Eq.~(\ref{blk_sh_ratio}), have an interesting interpretation \cite 
{Bluhm:2011xu, Dusling:2011fd}. Indeed, the change of $\zeta/\eta$ 
with $c_s^2$, for the linear ansatz, is the same as that found in a 
weakly coupled theory \cite{Weinberg, Arnold:2006fz}. On the other 
hand, the quadratic ansatz leads to a qualitative behaviour similar 
to that of the strongly coupled theories \cite{Buchel:2007mf}. 
Moreover, for quadratic ansatz, the result in Eq.~(\ref 
{blk_sh_ratio}) is exactly the same as the lower bound for 
$\zeta/\eta$ found in Ref.~\cite {Buchel:2007mf}. This is a very 
interesting and quite intriguing result.


\subsection{Case 3: $m=0$, $\mu\ne0$}

In this case, the non-vanishing dissipative quantities are the shear
stress tensor and the dissipative charge current. Considering the
$z\to 0$ limits of Eqs.~(\ref{en_den})-(\ref {num_den}) and
substituting in Eqs.~(\ref{evol_e})-(\ref {evol_n}), we get
\begin{align}
\dot\beta &= \frac{\beta}{3}\theta + {\cal O}(\delta^2), \quad
\dot\alpha = {\cal O}(\delta^2), \label{beta_alpha_dot_ch}\\
\nabla^\mu\beta &= - \beta\dot u^\mu + \frac{n}{\epsilon+P}\nabla^\mu\alpha
- \frac{\beta}{\epsilon+P}\Delta^\mu_\mu\partial_\nu\pi^{\mu\nu}. \label{beta_nab_ch}
\end{align}
Using the above relations, Eqs.~(\ref{first_ord_corr_L}) and (\ref
{first_ord_corr_Q}) becomes
\begin{align}
\delta f_L &= \frac{\beta\tau_0}{u\cdot p}\, f_0\left[ \left\{\frac{n(u\cdot p)}{\beta(\epsilon+P)}
-\frac{1}{\beta}\right\}p^\mu\nabla_\mu\alpha + p^\mu p^\nu\sigma_{\mu\nu} \right], \label{FOC_sh_ch_L}\\
\delta f_Q &= \frac{\beta\tau_0}{T}\, f_0\left[ \left\{\frac{n(u\cdot p)}{\beta(\epsilon+P)}
-\frac{1}{\beta}\right\}p^\mu\nabla_\mu\alpha + p^\mu p^\nu\sigma_{\mu\nu} \right], \label{FOC_sh_ch_Q}
\end{align}
up to first order in gradients.

Substituting Eqs.~(\ref{FOC_sh_ch_L}) and (\ref{FOC_sh_ch_Q}) in
Eqs.~(\ref{shear}) and (\ref{ch_curr}), we obtain the Navier-Stokes
equations for the shear stress tensor and   dissipative charge current,  as
\begin{align}
\pi^{\mu\nu} &= 2\, \tau_0\beta_\pi\, \sigma^{\mu\nu} , \label{sh_FOE}\\
n^\mu &= \tau_0\beta_n\, \nabla^\mu\alpha . \label{ch_FOE}
\end{align}
The first-order transport coefficient of the shear stress tensor for 
the two ansatzes, is obtained as
\begin{equation}\label{betapi_sh_ch}
\beta_\pi = \left\{ \begin{array}{rl}
\displaystyle \frac{1}{5}\left(\epsilon+P\right) &  \qquad \mbox{ (linear ansatz), } \\
\displaystyle \epsilon+P & \qquad \mbox{ (quadratic ansatz),} \end{array}\right.
\end{equation}
This is the same result as obtained in the Case 1 where $m=\mu=0$.

The first-order transport coefficient for the dissipative charge 
current is obtained as
\begin{equation}\label{betan_sh_ch}
\beta_n = \left\{ \begin{array}{rl}
\displaystyle \frac{\beta P}{3} - \frac{n^2T}{\epsilon+P} &  \qquad \mbox{ (linear ansatz), } \\
\displaystyle \beta P - \frac{4n^2T}{\epsilon+P} & \qquad \mbox{ (quadratic ansatz).} \end{array}\right.
\end{equation}
In this case, a comparison of Eqs.~(\ref{sh_FOE}) and (\ref{ch_FOE}) 
with Eqs.~(\ref{shear_NS}) and (\ref{charge_NS}) gives, $\eta= 
\tau_0\beta_\pi$ and $\kappa_n=\tau_0\beta_n$. Therefore, the ratio 
of the charge conductivity and the shear viscosity coefficients, 
$\kappa_n/\eta=\beta_n/\beta_\pi$, is independent of $\tau_0$. 
Moreover, the ratio of the heat conductivity and the shear 
viscosity, $\kappa_q/\eta=(\beta_n/\beta_\pi) [(\epsilon+P)/nT]^2$. 
In the small $\alpha$ limit, this ratio  is calculated as 
\begin{equation}\label{ht_sh_ratio}
\frac{\kappa_q}{\eta} = \left\{ \begin{array}{rl}
\displaystyle \frac{20}{3}\frac{T}{\mu^2} &  \qquad \mbox{ (linear ansatz), } \\
\displaystyle 4\frac{T}{\mu^2} & \qquad \mbox{ (quadratic ansatz).} \end{array}\right.
\end{equation}
The above equations are similar to the Wiedemann-Franz law \cite 
{Pitaevskii_Lifshitz, Son:2006em}. We also note that the ratio of 
the heat conductivity and shear viscosity exhibit an identical 
qualitative behaviour for the linear and quadratic ansatz. This 
property is in accordance with the previous results obtained for the 
strongly \cite{Son:2006em} and weakly coupled \cite 
{Danielewicz:1984ww} theories in the limit of small chemical 
potential.


\section{Conclusions and outlook}

We have considered the momentum dependent relaxation time 
$\tau_R(x,p)$ of the Boltzmann equation in the relaxation time 
approximation. The power law parametrization, $\tau_R(x,p)\sim p^a$, 
have been applied with the linear ($a=0$) and quadratic ($a=1$) 
ansatz. The main focus was to calculate the influence of the 
momentum dependent $\tau_R$ on the properties of the transport 
coefficients. We have employed the iterative Chapman-Enskog method 
to obtain the first-order solution of the Boltzmann equation with 
the momentum dependent relaxation time. We then derived expressions 
for transport coefficients such as the shear and bulk viscosity as 
well as the charge and heat conductivity.

We have shown that the first viscous correction to the distribution 
function derived using the linear and quadratic ansatz leads to 
identical expressions as those obtained from the Chapman-Enskog  and 
Grad's 14-moment method, respectively. We also demonstrated,  that 
the ratios of transport coefficients in these two cases corresponds 
to the weak and strong coupling regimes. In particular,  in the case 
of quadratic ansatz, we found that the ratio of the bulk and  shear 
viscosity  is exactly the same as the lower limit obtained for a 
system of a strongly coupled gauge theory plasma \cite
{Buchel:2007mf}. We also found, that in the limit of small chemical 
potential, the ratio of the heat conductivity to shear viscosity,  
for linear and quadratic ansatz, has a similar qualitative behavior, 
which is in agreement with the previous results obtained in the weak 
and strong coupling regimes.

Although we have considered a system of Boltzmann gas of single 
species to reduce cumbersome calculations, the present treatment can 
be rather easily generalized to a more complex system. We are also 
considering the extension of our results  to the second-order 
dissipative hydrodynamic equations with the quadratic ansatz on the 
momentum dependent relaxation time to compare the transport 
coefficients with those obtained by  using the moment method.


\begin{acknowledgements}
A.J. was supported by the Frankfurt Institute for Advanced Studies 
(FIAS). The work of B.F. was supported in part by the Extreme Matter 
Institute EMMI. K.R. acknowledges support by the Polish Science 
Foundation (NCN), under Maestro grant DEC-2013/10/A/ST2/00106, and 
of the U.S. Department of Energy under Grant No. DE-FG02- 05ER41367. 
K.R. also acknowledges discussion with S. Bass, M. Bluhm and M. 
Nahrgang.
\end{acknowledgements}

\end{document}